\documentclass[a4paper,11pt]{article}
\usepackage{amsmath}
\usepackage{amssymb}
\usepackage{bm}
\RequirePackage[numbers,sort&compress]{natbib}

\def\be{\begin{equation}}
\def\ee{\end{equation}}
\def\ba{\begin{eqnarray}}
\def\ea{\end{eqnarray}}
\def\onehalf{{\textstyle{\frac{1}{2}}}}

\def\omegaw{{\stackrel{\mbox{\tiny$\bullet$}}{\omega}}{}}

\def\jw{{\stackrel{\mbox{\tiny$~\bullet$}}{\jmath}}{}}

\def\Tw{{\stackrel{\mbox{\tiny$\bullet$}}{T}}{}}

\def\Kw{{\stackrel{\mbox{\tiny$\;\bullet$}}{K}}{}}
\def\Aw{{\stackrel{\mbox{\tiny$\bullet$}}{\omega}}{}}

\def\sw{{\stackrel{\mbox{\tiny$~\bullet$}}{S}}{}}

{}

\def\Abol{{\stackrel{\mbox{\tiny$\circ$}}{\omega}}{}}
\def\Gbol{{\stackrel{\mbox{\tiny$\,\circ$}}{G}}{}}

\def\Rbol{{\stackrel{\mbox{\tiny$~\circ$}}{R}}{}}

\def\omegabol{{\stackrel{\mbox{\tiny$\circ$}}{\omega}}{}}

\renewcommand{\thefootnote}{\fnsymbol{footnote}}

\begin{document}

\begin{center}
{\Large \bf An Improved Framework for Quantum Gravity}
\vskip 0.5cm
{\bf J. G. Pereira$^a$\footnote{Email: jg.pereira@unesp.br}, D. F. L\'opez$^{b}$\footnote{Email: dlopez@liceoingles.edu.co}}
\vskip 0.2cm
$^a$ {\it Instituto de F\'{\i}sica Te\'orica, 
Universidade Estadual Paulista (UNESP) \\
S\~ao Paulo, Brazil}
\vskip 0.2cm
$^b$ {\it Fundaci\'on Liceo Ingl\'es (FLI),
Departamento de Ciencias \\Pereira, Colombia} 
\end{center}


\vskip 0.3cm
\begin{quote}
{\bf Abstract.} {\small General relativity has two fundamental problems that render it unsuitable for tackling the gravitational field's quantization. The first problem is the lack of a genuine gravitational variable representing gravitation only, inertial forces apart. The second problem is its incompatibility with quantum mechanics, a problem inherited from the more fundamental conflict of special relativity with quantum mechanics. A procedure to overcome these difficulties is outlined, which amounts to replacing general relativity with its teleparallel equivalent and the Poincar\'e--invariant special relativity with the de Sitter-invariant special relativity. Those replacements give rise to the de Sitter-modified teleparallel gravity, which does not have the two mentioned problems. It can thus be considered an improved alternative approach to quantum gravity.
}
\end{quote}

\renewcommand{\thefootnote}{\arabic{footnote}}
\setcounter{footnote}{0}

\section{Introduction}
\label{intro}

Notwithstanding the success of Einstein's general relativity in describing a great many gravitational phenomena, it has encountered some difficulties related to its application to galactic, extragalactic, and cosmic scales. One of the most emblematic challenges is that Einstein's equation does not have a solution for a universe with accelerated expansion, as observations made in the last twenty years have indicated \cite{obs1,obs2,obs3}. One possibility is that general relativity fails to describe the gravitational interaction at the cosmic scale. If it does not fail, there might exist an elusive exotic fluid, generically called dark energy, which would be responsible for the observed accelerated expansion of the universe.

The difficulties mentioned above are not the only ones faced by general relativity: there is also the old question of quantizing the gravitational field \cite{Ash}. In principle, there are {\em two obstacles} that preclude general relativity from tackling the quantum--gravity problem. The first one refers to the gravitational variable of general relativity. It is well--known that general relativity is fundamentally grounded on the equivalence principle \cite{weinberg}, whose strong version establishes the {\it local} equivalence between gravitation and inertial effects. As a consequence of this approach, the spin connection of general relativity turns out to represent, in addition to gravitation, the inertial effects present in the frame used to describe the physical phenomenon.

That this is the case can be seen by observing that, in a locally inertial frame, in which inertial effects precisely compensate for gravitation, that connection vanishes, and gravitation becomes locally undetectable. Therefore, the spin connection of general relativity is not a genuine gravitational variable in the usual sense of classical field theory: it vanishes at a point where there is a non--vanishing gravitational field. This inertia--dependence of the spin connection, which is a consequence of the geometric character of general relativity, poses a severe difficulty to any spin connection approach to quantizing gravity because, together with gravitation, inertial forces would also be quantized. Given the divergent asymptotic behavior of the inertial forces, which is at odds with physical fields' behavior, one should expect consistency problems to turn up.

The second obstacle that precludes general relativity from tackling the quantum--gravity problem is its incompatibility with quantum mechanics. This problem stems actually from the more fundamental conflict of special relativity with quantum mechanics. One of the main points is the existence of a Lorentz invariant length at the Planck scale, given by the Planck length. Considering that the Lorentz subgroup of Poincar\'e forbids the existence of an invariant length, the Poincar\'e--invariant Einstein special relativity fails at the Planck scale \cite{1}. On account of the strong equivalence principle, if special relativity is inconsistent with quantum mechanics for energies comparable to the Planck energy, general relativity will be inconsistent either.

The purpose of the present paper is to discuss a procedure to overcome the two obstacles that render general relativity unsuitable for quantizing gravitation, giving rise to an improved framework to approach the quantum--gravity problem.

\section{Towards a Genuine Gravitational Variable}
\label{GenuineVar}

We are going to use the Greek alphabet ($\mu, \nu, \rho, \dots$) to denote spacetime indices, and the Latin alphabet ($a, b, c, \dots$) to denote tangent space (algebraic) indices. The metric of the tangent Minkowski space is assumed to be $\eta_{ab} = \mbox{diag}(1, -1, -1, -1)$. Quantities belonging to general relativity will be denoted with an over circle ``$^\circ$'', whereas quantities belonging to teleparallel gravity will be denoted with an over bullet ``$^\bullet$''.

\subsection{Special Relativity Spin Connection}

In special relativity, the equation of motion of a free particle, in an inertial frame, is written as
\be
\frac{d u'^a}{ds} = 0.
\label{FreeEq}
\ee
The corresponding equation of motion in a non-inercial frame is obtained by performing a local Lorentz transformation $\Lambda^a{}_c(x^\mu)$. Under such transformation, the anholonomic four--velocity changes according to
\be
u^a = \Lambda^a{}_c(x^\mu) u'^c.
\ee
Using the identity $\Lambda^a{}_b(x^\mu) \Lambda_e{}^b(x^\mu) = \delta^a_e$, the inverse transformation is found to be
\be
u'^a = \Lambda_e{}^a(x^\mu) u^e.
\ee
Substituting in the equation of motion (\ref{FreeEq}), we obtain \cite{TeleBook}
\be
\frac{d u^a}{ds} + \omegaw^a_{\ b \mu} u^b = 0,
\label{TGfEq0}
\ee
where
\be
\omegaw^a_{\ b \mu} = \Lambda^a_{\ c}(x) \partial_\mu \Lambda_b^{\ c}(x)
\label{InerConn0}
\ee
is the spin connection that represents the inertial forces present in the Lorentz--rotated frame.

\subsection{General Relativity Spin Connection}

In the tetrad formulation of general relativity, Einstein's equation is written as
\be
\Rbol^a{}_\mu - {\onehalf} h^a{}_\mu \Rbol = - \frac{8 \pi G}{c^4}\, {\mathcal T}^a{}_\mu,
\ee
where $\Rbol^a{}_\mu$ and $\Rbol$ are, respectively, the Ricci and the scalar curvatures, $h^a{}_\mu$ is the tetrad field, and ${\mathcal T}^a{}_\mu$ is the source energy--momentum current. The Riemann curvature is written as
\be
\Rbol^a{}_{b \mu \nu} = \partial_\nu \omegabol^a{}_{b \mu} - \partial_\mu \omegabol^a{}_{b \nu} 
+ \omegabol^a{}_{c \nu} \omegabol^c{}_{b \mu} - \omegabol^a{}_{c \mu} \omegabol^c{}_{b \nu}
\label{Riemann}
\ee
with $\omegabol^a{}_{b \mu}$ the Levi--Civita spin connection --- the spin connection of general relativity. In this theory, the equation of motion of a spinless particle is given by the geodesic equation
\be
\frac{d u^a}{ds} + \omegabol^a{}_{b \mu} u^b = 0.
\label{GeoEq}
\ee
The vanishing of the right--hand side means that there is no gravitational force in general relativity. Instead, the gravitational interaction is geometrized: gravitation produces a curvature in spacetime, and the particle's trajectories are given by the geodesics of the curved spacetime --- lines that follow the spacetime curvature.

The Levi--Civita spin connection plays a central role in the theory of Lorentz connections. Given a general Lorentz connection $\omega^a{}_{b \mu}$, it can be decomposed in the form \cite{KoNo}
\be
\omega^a{}_{b \mu} = \omegabol^a{}_{b \mu} + K^a{}_{b \mu}
\ee
where $K^a{}_{b \mu}$ is the contortion of the connection $\omega^a{}_{b \mu}$. In the specific case of the inertial spin connection $\omegaw^a{}_{b \mu}$, given by Eq.~(\ref{InerConn0}), the above decomposition assumes the form
\be
\omegabol^a{}_{b \mu} = \omegaw^a{}_{b \mu} - \Kw^a{}_{b \mu}.
\label{Decompo1}
\ee
This decomposition corresponds to a separation between inertial forces and gravitation: the spin connection $\Abol^a{}_{b \mu}$, in which inertial effects and gravitational are mixed, splits into the purely inertial piece $\Aw^a{}_{b \mu}$ and the purely gravitational piece $\Kw^a{}_{bc}$. In a locally inertial frame, where inertial effects precisely compensate for gravitation, the connection $\Abol^a{}_{b \rho}$ vanishes, and Eq.~(\ref{Decompo1}) assumes the form
\be
\Aw^a{}_{b \mu} \doteq \Kw^a{}_{b \mu},
\label{I=G}
\ee
which explicitly shows the local equivalence between inertial effects and gravitation, as established by the strong equivalence principle. Considering that the spin connection $\omegabol^{a}{}_{b \mu}$ vanishes at a point where there is a non--vanishing gravitational field, it cannot be considered a genuine gravitational variable in the usual sense of classical field theory.

\subsection{Teleparallel Gravity Spin Connection}

Substituting the decomposition (\ref{Decompo1}) into the geodesic equation (\ref{GeoEq}), we obtain the teleparallel force equation \cite{TeleBook}
\be
\frac{d u^a}{ds} + \omegaw^a{}_{b \mu} u^b = \Kw^a{}_{b \mu} u^b \, .
\label{TGfEq6}
\ee
In this equation, whereas the inertial effects $\omegaw^a{}_{b \mu}$ remain geometrized (in the sense of general relativity) in the left--hand side of the equation of motion, the contortion tensor $\Kw^a{}_{b \mu}$ plays the role of gravitational force in the right--hand side of the equation of motion. In the absence of gravitation, contortion vanishes and the force equation (\ref{TGfEq6}) reduces to the special relativity equation of motion (see Eq.~(\ref{TGfEq0}))
\be
\frac{d u^a}{ds} + \omegaw^a{}_{b \mu} u^b = 0.
\label{TGfEq8}
\ee
In this way, we see that the spin connection of teleparallel gravity is the very same spin connection that represents inertial effects in special relativity.

Although equivalent to general relativity, teleparallel gravity is an entirely different theory from the conceptual viewpoint. For example, in contrast to general relativity, teleparallel gravity has the structure of a gauge theory for the translation group \cite{gaugeTG}. The gravitational field in this theory is consequently represented by a gauge potential assuming values in the Lie algebra of the translation group,
\be
B_\mu = B^a{}_{\mu} \partial_a,
\ee
with $\partial_a$ the generators of infinitesimal translations. In the class of frames in which the inertial forces vanish, called proper frames,\footnote{The class of proper frames is the generalization, for the presence of gravitation, of the class of inertial frames.} the gauge (translational) coupling prescription is written as
\be
\partial_\mu \to D_\mu = \partial_\mu + B'^a{}_{\mu} \partial_a.
\label{tgcpres0}
\ee
Introducing the tetrad
\be
h'^a{}_\mu = \partial_\mu x'^a + B'^a{}_{\mu},
\ee
the coupling prescription can be rewritten in the form
\be
\partial_\mu \to D_\mu = h'^a{}_\mu \partial_a.
\label{tgcpres0Bis}
\ee

We consider now a local Lorentz transformation $\Lambda^a{}_{c}(x^\mu)$, which relates the class of proper frames to a general class of frames:
\be
x^a = \Lambda^a{}_b(x^\mu) x'^b \qquad \mbox{and} \qquad B^a{}_\mu = \Lambda^a{}_b(x^\mu) B'^b{}_\mu.
\ee
The inverse transformations are
\be
x'^a = \Lambda_e{}^a(x^\mu) x^e \qquad \mbox{and} \qquad B'^a{}_\mu = \Lambda_e{}^a(x^\mu) B^e{}_\mu.
\ee
Substituting into the coupling prescription (\ref{tgcpres0}), it assumes the form \cite{illus}\footnote{Note that the base members $\partial_a$ of the translational Lie algebra are not transformed.}
\be
\partial_\mu \to D_\mu = \partial_\mu + \omegaw^a{}_{b \mu} x^b \partial_a + 
B^a{}_{\mu} \partial_a.
\label{tgcpres}
\ee
This coupling prescription can be rewritten in the form
\be
\partial_\mu \to D_\mu = h^a{}_\mu \partial_a,
\label{tgcpresbis}
\ee
where
\be
h^a{}_\mu = \partial_\mu x^a + \omegaw^a{}_{b \mu} x^b + B^a{}_{\mu} 
\label{TeleTetra}
\ee
is the Lorentz-rotated frame, with~$\omegaw^a{}_{b \mu}$ the inertial spin connection (\ref{InerConn0}), which~represents the inertial forces present in the new frame $h^a{}_\mu$.

Like in all gauge theories, the commutator
\be
[D_\mu, D_\nu] = \Tw^a{}_{\mu \nu} \partial_a
\ee
yields the field strength of teleparallel gravity
\be
\Tw^a{}_{\mu \nu} = \partial_\mu B^a{}_\nu - \partial_\nu B^a{}_\mu +
\omegaw^a{}_{b \mu} B^b{}_\nu - \omegaw^a{}_{b \nu} B^b{}_\mu.
\label{TGfe}
\ee
Since the field strength of the trivial tetrad $e^a{}_\mu = \partial_\mu x^a + \omegaw^a{}_{b \mu} x^b$ vanishes, the field strength (\ref{TGfe}) is equivalent to the torsion 2--form
\be
\Tw^a{}_{\mu \nu} = \partial_\mu h^a{}_\nu - \partial_\nu h^a{}_\mu +
\omegaw^a{}_{b \mu} h^b{}_\nu - \omegaw^a{}_{b \nu} h^b{}_\mu.
\label{TeleTorsion}
\ee

\subsection{A Genuine Gravitational Variable}

Differently from general relativity, in which gravitation and inertial forces are both included in the Levi--Civita spin connection $\omegabol^a{}_{b \mu}$, in teleparallel gravity, the translational gauge potential $B^a{}_{\mu}$ represents the gravitational field, whereas the spin connection $\omegaw^a{}_{b \mu}$ represents inertial forces. We can then say that $B^a{}_{\mu}$ is a truly gravitational variable because it represents gravitation only, not inertial effects. As a translational connection, $B^a{}_{\mu}$ does not vanish in a locally inertial frame, being more akin to the usual classical fields concept. For these reasons, it is the natural variable to be quantized in any approach to quantum gravity. Furthermore, considering that teleparallel gravity has the geometric structure of a gauge theory, it is possible to use the well--known quantization techniques of internal gauge theories, including loop quantization \cite{loop1,loop2}, to address the quantum--gravity problem.

The inertia--dependence of the Levi--Civita spin connection is responsible for some peculiar properties of general relativity. For example, any attempt to obtain an expression for the energy--momentum density of gravity yields a complex that includes the energy--momentum density of both gravitation and inertial effects. Owing to the non--tensorial nature of inertial effects, any energy--momentum complex in this theory will be a pseudo--tensor. The same happens to the gravitational action written for solutions to Einstein's equation, which diverges when integrated over the whole spacetime \cite{York}. Considering that physical fields vanish at infinity, it is clear that inertial effects cause those divergences. It is then necessary to use a regularization process to eliminate the inertial forces and obtain a finite action \cite{GH}.

On the other hand, since inertial forces and gravitation are described by different variables in teleparallel gravity, it~turns out that it is possible to define energy--momentum density for gravitation only, to the exclusion of inertial forces. As expected, such energy--momentum density is a tensorial quantity \cite{PRL}. Analogously, it is possible to write down a gravitational action in which no inertial forces are present. Likewise, such action is naturally finite \cite{ActionRenorm}. In teleparallel gravity, each class of Lorentz frames is connected to a specific inertial spin connection. For example, the class of proper frames $h'^a{}_\mu$ is connected to a vanishing inertial spin connection $\omegaw'^a{}_{b \mu}=0$. Provided the appropriate spin connection for each frame is used, the computation of physical quantities will always be the physically relevant finite result \cite{Regularizing}.

\section{The de Sitter--Invariant Special Relativity}
\label{GraviQuanta}

For energies comparable to the Planck energy, quantum mechanics predicts the existence of a Lorentz invariant length, given by the Planck length. Since Einstein's special relativity does not allow invariant lengths, it turns out to be inconsistent with quantum mechanics. For this reason, there exists a widespread belief that the Lorentz symmetry should be broken down to allow the existence of an invariant length \cite{2,3,4,5}. However, this is not necessarily true. A less radical alternative, which preserves the Lorentz symmetry, is arguably to assume that, instead of governed by the Poincar\'e group, the spacetime kinematics be governed by the de Sitter group. Such assumption implies to replace the Poincar\'e--invariant Einstein special relativity with a {\em de Sitter invariant special relativity} \cite{dSsr0PRE}.

To understand the rationale behind this replacement, let us note that Lorentz is a subgroup of both Poincar\'e and de Sitter groups. Despite this fact, whereas the Poincar\'e group does not allow the existence of invariant lengths, the de Sitter group allows the existence of the particular length $l$ that defines the cosmological term
\be
\Lambda = 3 / l^{2},
\ee
which is known as the de Sitter length--parameter (or pseudo--radius). Considering that the cosmological term $\Lambda$ --- which represents the de Sitter spacetime's sectional curvature --- is invariant under Lorentz transformations, the de Sitter length--parameter $l$ must also be invariant.\footnote{As a matter of fact, the Poincar\'e group also admits the presence of the same invariant length--parameter. However, this is not immediately visible because, in the Minkowski spacetime, this length--parameter is infinite on account of the vanishing cosmological term.}

Suppose now we want the Planck length $l_P$ to be invariant under Lorentz transformations. In that case, it must represent the pseudo--radius of spacetime at the Planck scale, which will be a de Sitter spacetime with the Planck cosmological term $\Lambda_P = 3 / l_P^{2}$ \cite{grf}. In the de Sitter--invariant special relativity, therefore, the existence of a particular invariant length--parameter does not clash with the Lorentz invariance, which remains a symmetry at all energy scales. Consequently, the de Sitter--invariant special relativity turns out to be consistent with quantum mechanics \cite{paperDE}.

The de Sitter-invariant special relativity can be interpreted as a generalization of Einstein's special relativity for energies comparable to the Planck energy. Accordingly, it~might produce deviations from special relativity for energies near the Planck energy or at the large scale of the universe, that is, for galactic, extra-galactic, and~cosmic~scales. Due to the small energies and short distances involved in the solar system, no deviations are expected to occur in this case.

\section{The de Sitter--Modified General Relativity}
\label{dSmGR}

Considering that any change in special relativity will produce concomitant changes in general relativity, we review in this section the main properties of the de Sitter-modified general relativity, a gravitational theory consistent with de Sitter-invariant special relativity. The ensuing gravitational field equation will be referred to as the {\em de Sitter--modified Einstein equation}.

\subsection{de Sitter as a Homogeneous Space}
\label{dSHomoSpace}

As is well-known, Minkowski is  homogeneous under spacetime translations. This means that all points of Minkowski spacetime are equivalent under spacetime translations, whose generators are written as
\be
P_\rho = \delta^\alpha_\rho \partial_\alpha \, ,
\label{transiM}
\ee
with $\delta^\alpha_\rho$ the Killing vectors of spacetime translations.

Like Minkowski, the de Sitter spacetime is also homogeneous, but its homogeneity is deeply different from Minkowski's. In fact, all points of the de Sitter spacetime are equivalent under the so--called de Sitter ``translations'', whose generators are written as
\be
\Pi_\rho = \xi^\alpha_\rho \partial_\alpha,
\label{Pi}
\ee
with $\xi^\alpha_\rho$ the corresponding Killing vectors. The explicit form of the Killing vectors, as well as of the generators $\Pi_\rho$, depend on the coordinate system used to write them.

In stereographic coordinates $\{x^\mu\}$ \cite{Gursey}, the Killing vectors of the de Sitter ``translations'' have the form \cite{dSgeod}
\be
\xi^\alpha_\rho = \delta^\alpha_\rho - \big(1/4l^2 \big) \bar \delta^\alpha_\rho,
\label{KilVecdStrans}
\ee
where $\delta^\alpha_\rho$ are the Killing vectors of translations and
\be
\bar \delta^\alpha_\rho = 2 \eta_{\rho \nu} \, x^\nu x^\alpha - 
\sigma^2 \delta_\rho^\alpha
\ee
are the Killing vectors of the proper (or local) conformal transformations. Substituting in Eq.~(\ref{Pi}), the de Sitter ``translation'' generators assume the form
\be
{\Pi}_\rho = {P}_\rho - \big(1/4l^2 \big) {K}_\rho
\label{pi3}
\ee
where
\be
{P}_\rho = \partial/ \partial x^\rho \quad \mbox{and} \quad
{K}_\rho = \big(2 \eta_{\rho \nu} \, x^\nu x^\mu - \sigma^2 \delta_{\rho}^{\mu} \big) 
\partial/ \partial x^\mu
\label{TransGenerators}
\ee
are, respectively, the translation and the proper conformal generators \cite{coleman}. Comparing Eqs.~(\ref{transiM}) and (\ref{pi3}) we see that, whereas Minkowski is homogeneous under translations, de Sitter is homogeneous under a combination of translations and proper conformal transformations. 

\subsection{Conservation Law of Source Fields}
\label{MCLaw}

In general relativity, the strong equivalence principle says that the physical laws reduce locally to those of special relativity. In this case, all solutions to Einstein equation will be spacetimes that reduce locally to Minkowski, where the local kinematics governed by the Poincar\'e group takes place. On the other hand, when the de Sitter--invariant special relativity replaces the Poincar\'e--invariant special relativity, all solutions to the {\em de Sitter--modified Einstein equation} will be spacetimes that reduce locally to de Sitter, where the local kinematics takes place.

Let us consider the action integral of a general source (or matter) field
\be
{\mathcal S}_m = \frac{1}{c} \int {\mathcal L}_m \sqrt{-g} \, d^4x,
\label{Am}
\ee
with ${\mathcal L}_m$ the Lagrangian density. Invariance of this action under general coordinate transformations yields --- through Noether's theorem --- the covariant conservation of the current appearing in the right--hand side of Einstein's equation. As local transformations, diffeomorphisms can detect the local structure of spacetime. For example, in the case of general relativity, where all solutions to Einstein's equations are spacetimes that reduce locally to Minkowski, a diffeomorphism is defined by
\be
\delta_P x^\mu \equiv \varepsilon^\rho(x) P_\rho x^\mu =
\delta^\mu_\alpha \varepsilon^\alpha(x),
\label{OrDiff}
\ee
with $\delta^\mu_\alpha$ the Killing vectors of translations --- which are the transformations that define Minkowski's homogeneity. The corresponding transformation of the metric tensor is
\be
\delta_P g_{\mu \nu} = - \delta^\alpha_\nu \nabla_\mu \varepsilon_\alpha(x) -
\delta^\alpha_\mu \nabla_\nu \varepsilon_\alpha(x).
\label{Orddeltag}
\ee
Variation of ${\mathcal S}_m$ with respect to this metric transformation gives
\be
\delta {\mathcal S}_m = -\,  \frac{1}{2 c} \int  {\mathcal T}^{\mu \nu} \delta_P g_{\mu \nu} \, \sqrt{-g} \, d^4x, 
\label{OrdiSvar4}
\ee
where
\be
{\mathcal T}^{\mu \nu} = \frac{2}{\sqrt{-g}} \, \frac{\delta {\mathcal L}_m}{\delta g_{\mu \nu}}
\label{syem}
\ee
is the symmetric energy--momentum tensor. Substituting the metric transformation (\ref{Orddeltag}) and transferring the Killing vectors appearing in $\delta_P g_{\mu \nu}$ to the current definition, the action variation assumes the form
\be
\delta {\mathcal S}_m = \frac{1}{2 c} \int \big(\delta^\mu_\alpha {\mathcal T}^{\alpha \nu} \big) \delta g_{\mu \nu} \, \sqrt{-g} \, d^4x, 
\label{OrdiSvar4bis}
\ee
where $\delta^\mu_\alpha {\mathcal T}^{\alpha \nu} \equiv {\mathcal T}^{\mu \nu}$ is the current definition, and $\delta g_{\rho \mu}$ is an arbitrary metric transformation, that is, a transformation that does not depend on the Killing vectors defining the local homogeneity of spacetime. Variation (\ref{OrdiSvar4bis}) is a fundamental relation in that it defines the form of the conserved current. For this reason, it is the variation to be used in the variational principle for obtaining the gravitational field equations. It is also the variation to be used in Noether's theorem for obtaining the covariant conservation law of the source fields. In this case, substituting the arbitrary metric transformation
\be
\delta g_{\mu \nu}  = - \nabla_\mu \varepsilon_\nu(x) -
\nabla_\nu \varepsilon_\mu(x),
\label{GeneMetTrans}
\ee
and following the standard procedure \cite{LanLif}, the invariance of ${\mathcal S}_m$ yields the covariant conservation law
\be
\nabla_\nu \big(\delta^\mu_\alpha {\mathcal T}^{\alpha \nu} \big) \equiv
 \nabla_\nu {\mathcal T}^{\mu \nu} = 0,
\label{Tcon}
\ee
with the conserved current given by the projection of the energy-momentum tensor along the Killing vectors of ordinary translations, which in this case is the energy--momentum current itself.

We consider now the case of spacetimes that reduce locally to de Sitter, in which a diffeomorphism is defined by
\be
\delta_{\Pi} x^\mu \equiv \varepsilon^\rho(x) \Pi_\rho x^\mu = 
\xi^{\mu}_{\alpha} \, \varepsilon^{\alpha}(x),
\label{dStrans}
\ee
with $\xi^{\mu}_{\alpha}$ the Killing vectors of the de Sitter ``translations'' --- which are the transformations that define de Sitter's homogeneity. The corresponding transformation of the metric tensor is \cite{dSgeod}
\be
\delta_{\Pi} g_{\mu \nu} = -\, \xi^\alpha_\mu \nabla_\nu \varepsilon_\alpha(x) -
\xi^\alpha_\nu \nabla_\mu \varepsilon_\alpha(x) \, .
\label{dSmetricTrans}
\ee
Variation of the action (\ref{Am}) with respect to this metric transformation gives
\be
\delta {\mathcal S}_m = \frac{1}{2 c} \int {\mathcal T}^{\mu \nu} \delta_{\Pi} g_{\mu \nu} \, \sqrt{-g} \, d^4x
\label{Svar4}
\ee
with ${\mathcal T}^{\mu \nu}$ the symmetric energy--momentum tensor (\ref{syem}). Substituting the metric transformation (\ref{dSmetricTrans}) and transferring the Killing vectors appearing in $\delta_\Pi g_{\mu \nu}$ to the current definition, the action variation assumes the form\footnote{Since $\Pi^{\mu \nu}$ appears contracted with $\delta g_{\mu \nu}$, only its symmetric part will contribute to the gravitational field equations. From now on, it is implicitly understood that $\Pi^{\mu \nu}$ denotes the symmetric part of the tensor.}
\be
\delta {\mathcal S}_m = \frac{1}{2 c} \int \big(\xi^{\mu}_{\alpha} {\mathcal T}^{\alpha \nu} \big) \delta g_{\mu \nu} \, \sqrt{-g} \, d^4x,
\label{deltaA2}
\ee
where $\xi^{\mu}_{\alpha} {\mathcal T}^{\alpha \nu} \equiv \Pi^{\mu \nu}$ is the current definition, and $\delta g_{\rho \mu}$ is an arbitrary metric transformation, that is, a transformation that does not depend on the Killing vectors defining the local homogeneity of spacetime. Like in the case of locally--Minkowski spacetimes, variation (\ref{deltaA2}) is a fundamental relation in that it defines the form of the conserved current. For this reason, it is the variation to be used in the variational principle for obtaining the gravitational field equations. It is also the variation to be used in Noether's theorem for obtaining the covariant conservation law of the source fields. In this case, substituting the arbitrary metric transformation
\be
\delta g_{\mu \nu}  = - \nabla_\mu \varepsilon_\nu(x) -
\nabla_\nu \varepsilon_\mu(x),
\label{GeneMetTransds}
\ee
and following the standard procedure, the invariance of ${\mathcal S}_m$ yields the covariant conservation law
\be
\nabla_\nu \big(\xi^{\mu}_{\alpha} {\mathcal T}^{\alpha \nu} \big) \equiv
\nabla_\nu \Pi^{\mu \nu} = 0,
\label{UniCon}
\ee
with the conserved current $\Pi^{\mu \nu}$ given by the projection of the energy--momentum tensor along the Killing vectors of the de Sitter ``translations''. 

\subsection{The de Sitter--Modified Einstein Equation}

The Einstein--Hilbert action of the de Sitter--modified general relativity is similar to the action of ordinary general relativity:
\be
{\mathcal S} = - \frac{c^3}{16 \pi G} \int {\mathcal R} \, \sqrt{-g} \, d^4x.
\label{Sg}
\ee
However, there is a crucial difference. Considering that spacetime reduces locally to de Sitter, the scalar curvature ${\mathcal R}$, in this case, includes both the dynamical curvature of general relativity and the kinematic curvature of the underlying de Sitter spacetime. Under an arbitrary metric transformation $\delta g_{\mu \nu}$, the gravitational action ${\mathcal S}$ changes according to
\be
\delta {\mathcal S} = \frac{c^3}{16 \pi G} \int \big( {\mathcal R}^{\mu \nu} - \onehalf g^{\mu \nu} {\mathcal R} \big) \delta g_{\mu \nu} \, \sqrt{-g} \, d^4x.
\ee
The invariance of the total action $\delta{\mathcal S}_m + \delta {\mathcal S} = 0$, with $\delta{\mathcal S}_m$ given by Eq.~(\ref{deltaA2}), yields the de Sitter--modified Einstein equation \cite{paperDE}
\be
{\mathcal R}^{\mu \nu} - {\onehalf} g^{\mu \nu} {\mathcal R} = - \frac{8 \pi G}{c^4}\, \Pi^{\mu \nu}.
\label{NewEinstein}
\ee

By construction, all solutions to this equation are spacetimes that reduce locally to de Sitter, in which the local kinematics takes place. Considering that the de Sitter--invariant special relativity is consistent with quantum mechanics, the de Sitter--modified Einstein equation (\ref{NewEinstein}) will be likewise consistent. 
Another relevant point is that the cosmological term $\Lambda$ is now encoded in the local spacetime kinematics. As a consequence, it does not appear explicitly in the gravitational field equation. Accordingly, it is not constrained to be constant by the second Bianchi identity
\be
\nabla_\nu \big({\mathcal R}^{\mu \nu} - \onehalf g^{\mu \nu} {\mathcal R} \big) = 0. 
\label{BianciId}
\ee
This is a crucial property of the de Sitter--modified general relativity, which may have important consequences for gravitation, particularly cosmology \cite{ccc}.

\section{The de Sitter--Modified Einstein Equation in Stereographic Coordinates}

In stereographic coordinates, where the Killing vectors are given by Eq.~(\ref{KilVecdStrans}), the source of the field equation (\ref{NewEinstein}) splits according to
\be
\Pi^{\mu \nu} \equiv \xi^\mu_\alpha {\mathcal T}^{\alpha \nu} = {\mathcal T}^{\mu \nu} - (1/4 l^2) {\mathcal K}^{\mu \nu},
\label{Pi=T+K}
\ee
where 
\be
{\mathcal T}^{\mu \nu} \equiv \delta^\mu_\alpha {\mathcal T}^{\alpha \nu} \, \quad \mbox{and} \quad 
{\mathcal K}^{\mu \nu} \equiv \bar \delta^\mu_\alpha {\mathcal T}^{\alpha \nu} =
(2 \eta_{\alpha \beta} x^\beta x^\mu - \sigma^2 \delta^\mu_\alpha ) {\mathcal T}^{\alpha \nu}
\label{Kcurrent}
\ee
are, respectively, the symmetric energy--momentum current and the proper conformal current \cite{coleman}, with $\delta^\mu_\alpha$ and $\bar \delta^\mu_\alpha$ the Killing vectors of translations and proper conformal transformations. In these coordinates, the de Sitter--modified Einstein equation (\ref{NewEinstein}) assumes the form
\be
{\mathcal G}^{\mu \nu} \equiv {\mathcal R}^{\mu \nu} - \onehalf g^{\mu \nu} {\mathcal R} =
- \frac{8 \pi G}{c^4} \, \Big[{\mathcal T}^{\mu \nu} - 
(1/4 l^2) {\mathcal K}^{\mu \nu} \Big],
\label{NewEinsteinBis2}
\ee
with ${\mathcal G}^{\mu \nu}$ the Einstein tensor. In view of the source separation, the Einstein tensor ${\mathcal G}^{\mu \nu}$ can also be separated in the form\footnote{It should be noted that this separation can only be made in the Einstein tensor. It cannot be made in the metric, connection or curvature tensor.}  
\be
{\mathcal G}^{\mu \nu} = \Gbol^{\mu \nu} - 
\hat{G}^{\mu \nu} \, ,
\label{Gseparation}
\ee
where $\Gbol^{\mu \nu}$ is the Einstein tensor of the dynamical curvature of general relativity and $\hat{G}^{\mu \nu}$ is the Einstein tensor of the kinematic curvature of the background de Sitter spacetime. The minus sign of the decomposition comes from the minus sign of the source decomposition (\ref{Pi=T+K}). Substituting in Eq.~\eqref{NewEinstein}, we get
\be
\big(\Rbol^{\mu \nu} - {\onehalf} g^{\mu \nu} \Rbol \big) -
\big(\hat{R}^{\mu \nu} - {\onehalf} g^{\mu \nu} \hat{R}\big) = 
\frac{8 \pi G}{c^4} \Big[{\mathcal T}^{\mu \nu} - (1/4 l^2) {\mathcal K}^{\mu \nu} \Big].
\label{NewEinstein2}
\ee
Although written in a spacetime with metric $g^{\mu \nu}$, the Ricci and scalar curvatures of the local de Sitter spacetime can be written as
\be
\hat{R}^{\mu \nu} = \Lambda g^{\mu \nu}  \qquad \mbox{and} \qquad  
\hat{R} = 4 \Lambda.
\label{dSrelations}
\ee 
Since the metric $g^{\mu \nu}$ reduces locally to the de Sitter metric $\hat{g}^{\mu \nu}$, these tensors reduce locally to the Ricci and scalar curvatures of the background de Sitter spacetime. Using these relations, Eq.~\eqref{NewEinstein2} assumes the form
\be
\Rbol^{\mu \nu} - {\onehalf} g^{\mu \nu} \Rbol + g^{\mu \nu} \Lambda = -
\frac{8 \pi G}{c^4} \Big[{\mathcal T}^{\mu \nu} - (1/4 l^2) {\mathcal K}^{\mu \nu} \Big].
\label{NewEinstein3}
\ee
This is the de Sitter--modified Einstein equation in stereographic coordinates. Considering that the symmetric energy--momentum tensor ${\mathcal T}^{\mu \nu}$ is the source of the dynamic curvature of general relativity, {\em the proper conformal current ${\mathcal K}^{\mu \nu}$ shows up as the source of the kinematic de Sitter curvature}~\cite{OriginLambda}.

The field equation (\ref{NewEinstein3}) resembles the usual Einstein equation in the presence of a cosmological constant $\Lambda$. However, there is a crucial difference: the associated Bianchi identity, given by Eq.~(\ref{BianciId}), holds for the total Einstein tensor ${\mathcal G}^{\mu \nu}$, not for the general relativity Einstein tensor $\Gbol^{\mu \nu}$. In other words, it holds for the whole left--hand side of the field equation (\ref{NewEinstein3})
\be
\nabla_\nu \big(\Rbol^{\mu \nu} - {\onehalf} g^{\mu \nu} \Rbol - g^{\mu \nu} \Lambda\big) = 0,
\label{Bianchi}
\ee
where now the covariant derivative of the cosmological term $\Lambda$ is non--vanishing:
\be
\nabla_\nu \Lambda \equiv \partial_\nu \Lambda \neq 0.
\ee
The Bianchi identity \eqref{Bianchi} is consistent with the fact that the symmetric energy--momentum tensor is no longer covariantly conserved: $\nabla_\nu {\mathcal T}^{\mu \nu} \neq 0$. What is covariantly conserved now is the de Sitter current, $\nabla_\nu {\Pi}^{\mu \nu} = 0$, with ${\Pi}^{\mu \nu}$ a combination of the symmetric energy--momentum tensor and the proper conformal current, as given by Eq.~ (\ref{Pi=T+K}). 

It is important to emphasize that only in stereographic coordinates the cosmological term $\Lambda$ appears explicitly in the de Sitter--modified Einstein equation. For other coordinates, the curvature of the background de Sitter spacetime and the dynamic curvature of general relativity remain mixed in the Einstein tensor ${\mathcal G}^{\mu \nu}$, and the field equation cannot be recast in the form (\ref{NewEinstein3}). 

\section{The de Sitter--Modified Teleparallel Gravity}
\label{dSmTG}

In the tetrad formulation, the de Sitter--modified Einstein equation (\ref{NewEinstein3}) reads
\be
\Rbol^{a \nu} - {\onehalf} h^{a \nu} \Rbol + h^{a \nu} \Lambda = -
\frac{8 \pi G}{c^4}\, \Big[{\mathcal T}^{a \nu} - (1/4 l^2) {\mathcal K}^{a \nu} \Big].
\label{NewEinstein4}
\ee
The stereographic coordinate's main virtue is that it allows us to separate the dynamic curvature of gravitation from the local de Sitter spacetime's kinematic curvature. Considering that only the dynamic gravitational field should be quantized, this form of the field equation is quite convenient for quantization purposes. However, because general relativity is a geometric theory for gravitation, it does not have a genuine gravitational variable, which only represents gravitation. On the other hand, owing to its gauge structure, teleparallel gravity does have a genuine gravitational variable. A natural solution to the gravitational variable's problem is to use, instead of the de Sitter--modified Einstein equation (\ref{NewEinstein4}), the de Sitter--modified teleparallel field equation.

The first step to obtain the de Sitter--modified teleparallel field equation is to write down the teleparallel and the matter actions in a locally--de Sitter spacetime ($k = 8 \pi G/c^4$),
\be
{\mathcal S} = \frac{1}{4 k c} \int \Tw_{a \mu \nu} \, \sw^{a \mu \nu} \, h \, d^4x +
\frac{1}{c} \int {\mathcal L}_m \, h \, d^4x,
\ee
where $h = \det(h^a{}_\mu) = \sqrt{-g}$, $\Tw_{a \mu \nu}$ is the torsion, and $\sw^{a \mu \nu}$ is the superpotential. Then, one has just to follow the procedure used in Section~\ref{dSmGR} to obtain the de Sitter--modified Einstein's equation.

However, it is much simpler to obtain this equation by cosidering the teleparallel equivalent of the de Sitter--modified Einstein equation (\ref{NewEinstein4}). This can be done by using the well--known equivalence between the field equations of general relativity and teleparallel gravity, which is expressed by the identity \cite{TeleBook}
\be
\Rbol^{a \nu} - {\onehalf} h^{a \nu} \Rbol = 
\frac{1}{h} \Big[\partial_\mu \big(h \sw^{a \nu \mu} \big) -
k h \jw^{a \nu} \Big].
\label{EquiEqua}
\ee
Substituting this relation in the de Sitter--modified Einstein equation (\ref{NewEinstein4}), we obtain 
\be
\partial_\mu \big(h \sw^{a \nu \mu} \big) -
k h \jw^{a \nu} + h h^{a \nu} \Lambda = - k h \Big[{\mathcal T}^{a \nu} - (1/4 l^2) {\mathcal K}^{a \nu} \Big].
\label{dSMTFE}
\ee
This is the de Sitter--modified teleparallel field equation. At the classical level, the field equations (\ref{NewEinstein4}) and (\ref{dSMTFE}) are equivalent in the sense that they give the same physical results. At the quantum level, however, the latter has the advantage of bearing a genuine gravitation variable --- a variable that represents gravitation only --- being for this reason more appropriate to deal with the quantum gravity problem.

\section{Final Remarks}
\label{FR}

To highlight the main points, we summarize in this section the arguments leading to the de Sitter--modified teleparallel gravity as an alternative approach to quantum gravity. We begin by noting that, strictly speaking, general relativity is not a classical field theory but a geometric theory for the gravitational interaction. As a consequence of this approach, it lacks a variable that describes gravitation only, to the exclusion of inertial forces. For example, the Levi--Civita spin connection includes, in addition to gravitation, inertial forces present in the frame used to describe the gravitational phenomenon. On the other hand, in teleparallel gravity, inertial forces and gravitation are described by different variables. Whereas inertial forces are represented by the inertial spin connection $\omegaw^a{}_{b \mu}$, gravitation is represented by the translational gauge potential $B^a{}_\mu$. We can then say that a natural way to overcome this difficulty is to replace general relativity with the teleparallel equivalent of general relativity. The existence of a genuine gravitational variable in teleparallel gravity makes it a unique framework for dealing with the quantization of gravity.

However, the simple replacement of general relativity by its teleparallel equivalent is not enough to obtain a consistent approach. The problem is that teleparallel gravity, as well as general relativity, are inconsistent with quantum mechanics. Such inconsistency comes from the more fundamental inconsistency of special relativity, which holds locally in spacetime according to the strong equivalence principle. A solution to this inconsistency is arguably to replace the Poincar\'e--invariant Einstein special relativity with the de Sitter--invariant special relativity \cite{dSsr0PRE}. Considering that $\Lambda$ is invariant under local Lorentz transformations, the de Sitter length--parameter $l$, which relates to $\Lambda$ through $\Lambda = 3/l^2$, is also invariant. The de Sitter--invariant special relativity is, consequently, consistent with the Planck scale's physics, where the Lorentz invariant Planck--length plays a crucial role. 

Of course, if special relativity changes, all relativistic theories must change accordingly. In particular, teleparallel gravity changes to what has been called {\em de Sitter--modified teleparallel gravity}. In this theory, any solution to the de Sitter--modified teleparallel field equation is a spacetime that reduces locally to de Sitter. In addition to being consistent with quantum mechanics, it has a genuine gravitational variable, which describes gravitation only, not inertial forces. It constitutes, for these reasons, a consistent framework for approaching the quantum--gravity problem. It can be classified as a foundational approach to quantum gravity \cite{Lee3Roads}, standing apart from the usual superstring and loop quantum gravity approaches.

It is important to mention that there are other problems awaiting a better understanding, in addition to the two questions addressed in this paper. One of them is how to generalize the frame's notion to the quantum domain \cite{Qframe1, Qframe2}. Another one is whether the equivalence principles remain valid at the Planck scale \cite{FR4}. Putting it differently, one could wonder how the equivalence principles should be changed to meet the quantum requirements \cite{Qep}. Concerning this point, let us recall that Einstein's general relativity presupposes the equivalence principle's validity. Therefore, if the equivalence principle fails at the quantum level, general relativity would break down in that domain.

As a gauge theory, teleparallel gravity does not describe the gravitational interaction through a spacetime geometrization but as a gravitational force---a property shared with all other classical gauge theories. Like Newtonian gravity, it~was constructed to comply with the equivalence principles. However, in the absence of the principles, teleparallel gravity can still provide a consistent description of the gravitational interaction~\cite{AbsenceUniver, Einstein05}. If~the equivalent principles do not remain valid at the Planck scale, such feature can be considered an additional advantage of teleparallel gravity over general relativity for tackling the quantization of the gravitational~field.

\section*{Acknowledgments}
JGP thanks Conselho Nacional de Desenvolvimento Cient\'{\i}fico e Tecnol\'ogico, Brazil, for a research grant (Grant No. 304190/2017-9).


\end{document}